\documentclass[12pt]{article}
\usepackage{epsfig}

\renewcommand\vec[1]{\mathchoice{{\mathord{\mbox{\boldmath$\displaystyle #1$}}}}{{\mathord{\mbox{\boldmath$\textstyle #1$}}}}{{\mathord{\mbox{\boldmath$\scriptstyle #1$}}}}{{\mathord{\mbox{\boldmath$\scriptscriptstyle #1$}}}}}
\newcommand\sfrac[2]{{\textstyle\frac{#1}{#2}}}
\newcommand\xput[3]{\setbox1\hbox{\kern#1 \vtop{\kern#2\hbox{#3}}}\wd1=0pt \ht1=0pt \dp1=0pt \box1 \ifvmode\nointerlineskip\fi}

\begin{document}

\title{Orbital physics: glorious past, bright future}

\author{D. I. Khomskii\\[\medskipamount]
\normalsize II. Physikalisches Institut, Universit\"at zu K\"oln,\\[-\smallskipamount]
\normalsize Z\"ulpicher Str.\ 77,\\[-\smallskipamount]
\normalsize 50937 K\"oln, Germany}

\date{}

\maketitle

\xput{9.5cm}{-6.6cm}{{\it\footnotesize Dedicated to J.~B.~Goodenough}}
\xput{9.5cm}{-6.2cm}{{\it\footnotesize in honour of his 100th birthday}}

\begin{abstract}
\noindent
Transition metal (TM) compounds present a very big class of materials
with quite diverse properties. There are among them insulators, metals,
systems with insulator--metal transitions; most magnetic systems
are TM compounds; there are among them also (high-$T_c$) superconductors.
Their very rich properties are largely determined by the strong interplay
of different degrees of freedom: charge; spin; orbital; lattice.
Orbital effects play a very important role in these systems --- and not only in them!
The study of this field, initiated by Goodenough almost 70 years ago,
turned out to be very fruitful and produced a lot of important results.
In this short review I first discuss the basics of orbital physics,
summarize the main achievements in this big field, in which Goodenough
played a pivotal role and which are nowadays widely used to explain many
properties of TM compounds. In the main part of the text I discuss
novel developments and perspectives in orbital physics, which is still
a very active field of research, constantly producing new surprises.
\end{abstract}

\section{Introduction}
Magnetism is one of the first physical phenomena known to
mankind, from ancient Greece which gave us the very word
``magnetism'' --- although it was known (and used) even
earlier in ancient China. But it was only in the XX century, after the emergence
of quantum mechanics, that the real microscopic nature of magnetism
was understood.

\begin{figure}
\centering
\includegraphics{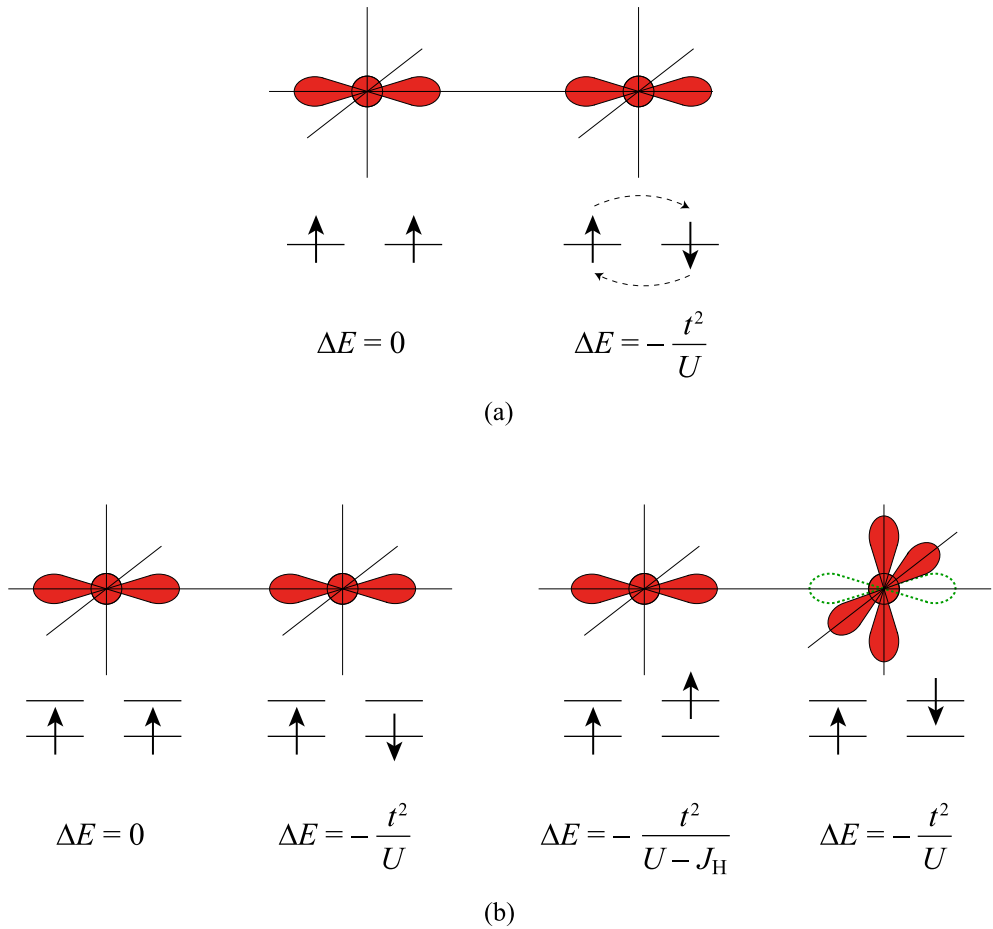}
\caption{\label{FIG:1}\footnotesize Orbital-dependent exchange interaction.
(a)~Nondegenerate orbitals. Virtual intersite hopping is allowed for antiparallel spins,
right, giving the energy gain $\Delta E = -2t^2/U$ (in the second order in~$t/U$), but
is forbidden by the Pauli principle for parallel spins, left. As a result we get the antiferromagnetic
exchange.
(b)~Exchange for two orbitals per site, with only diagonal hopping $t_{11}$, $t_{22}$,
with hopping between different orbital absent, $t_{12}=0$. The energy gain
due to virtual electron hopping is shown for each case. Due to Hund's rule
interatomic interaction $J_{\rm H}$ the third configuration --- different orbitals
and same spins --- is favoured due to virtual hopping of the elctron from the left
to the empty orbital on the right (shown by dashed line), giving ferromagnetic
exchange $\sim(t^2/U)(J_{\rm H}/U)$ (for $J_{\rm H}<U$).
}
\end{figure}

The present understanding of magnetism in most cases is that it is
due to, predominantly, spins of electrons, and to have strong
magnetic response one needs electrons localised at respective ions.
This is now the subject of a very rich field of correlated electrons,
initiated at the end of the 30s by Peierls \cite{Mott} (see
the detailed description of this history in Appendix A.\thinspace1 in \cite{KhomskiiBook}),
and developed later by Landau, Mott, Anderson and Hubbard~\cite{Landau1,Mott2,Anderson,Hubbard}.
There exist two types of states of electrons in solids: itinerant
electrons, well described by the band theory, and localised electrons ---
localised due to sufficiently strong electron--electron interactions.
In the simplest form the situation can be described by the Hubbard
model --- which looks deceptively simple but which hides a lot of
complications and surprises.  In the standard form it treats
electrons in a lattice made from ions with nondegenerate levels,
with, usually, the nearest-neighbour hoppings of electrons and with, in the simplest form, the on-site Coulomb repulsion
\begin{equation}
H = -t \sum_{\langle ij\rangle,\sigma} c_{i\sigma}^\dagger c_{j\sigma}^\dagger  + U\sum_i n_{i\uparrow}n_{i\downarrow}\,,
\qquad
n_{i\sigma}=c^\dagger_{i\sigma} c_{i\sigma}^{\vphantom{\dagger}}\,.
\label{eq:1}
\end{equation}
For one electron per site without the Coulomb (Hubbard) interaction,
$U=0$ or very small, the material in the band theory would contain
a half-filled band and would be a metal (we ignore here the possibility
of a Peierls distortion which can still make such system insulating).
However for strong enough interaction $U>W = 2zt$  (where $W$ is the
bandwidths and $z$ is the number of nearest neighbours) the electrons would be
localised one per site, and the system would be insulating~\cite{Mott}
--- what we now call Mott insulators. Such localised
electrons would also have localized spins, which would have a certain
exchange interaction, and at low enough temperatures there would
appear a long-range magnetic ordering. In simplest cases the exchange
(or superexchange) interaction of such localised electrons is
obtained in perturbation theory in $t/U \ll 1$, and it leads to the
antiferromagnetic Heisenberg exchange
\begin{equation}
H = J \sum_{\langle ij\rangle} \vec S_i\cdot \vec S_j\,,  \qquad   J = 2t^2/U \,.
\label{eq:2}
\end{equation}
Why this exchange is antiferromagnetic is illustrated in
Fig.~\ref{FIG:1}: in the second order in perturbation theory (in~$t/U$) one always
gains energy, but for electrons with parallel spins this process
of virtual hoppings is forbidden by the Pauli principle and we gain no
such energy. But for antiparallel spins this process is allowed, and
the corresponding energy gain stabilises such antiferromagnetic spin
ordering.

This simple treatment describes the basic physics of magnetism of
system with localised electrons. But in real materials
several very important extra factors have to be taken into account.
And it is here that John Goodenough came to the forefront, and in
effect he played a crucial role in making this simple model realistic,
and laid the groundwork for the real explanation of multitude of very
different magnetic phenomena. His contribution is ``formalised''
in the famous Goodenough--Kanamori, or, sometimes, Goodenough--Kanamori--Anderson (GKA) rules.  But John's
contribution is not only restricted to formulating these rules; he
demonstrated the power of this approach on numerous examples of
a multitude of real materials with very rich and often very puzzling
properties.

Such extra factors which one needs to take into account are of course
the specific crystal structure of a given material, but also its
orbital structure: transition metal compounds with their $d$-electrons
(but also rare earth and actinide systems), to which most of magnetic
materials belong, have not electrons on nondegenerate levels but
$d$-electrons with orbital moment $l=2$. Orbital degrees of freedom can
drastically alter the properties of a material, and their study
presents a very interesting and rich field of orbital physics, the
foundations of which were laid by Goodenough. This will be the topic
of the present paper. In this text I will first give a very short
summary of the basics and present well-established results, which is really
the glorious page of the very successful period of the development of
this field; but I will mostly discuss novel development and the
perspectives of this field, which is still very much alive
and which produces more and more surprises.  Necessarily my
discussion of this very big field will be rather qualitative; one can find more
detailed presentation of these topics in the rather
old, but still not obsolete review~\cite{KugelKhomskii}, and in more recent
reviews~\cite{Tokura,Khomskii1,Khaliullin,Streltsov,Khomskii2}. And of course one should mention
the famous book by Goodenough, {\it Magnetism and the chemical bond}~\cite{GoodenoughBook} ---
a wonderful book more than 50 (almost 60!)\ years old, which
still remains the bible in this field.  For me personally it is still
one of the main sources of information: although I myself already
wrote a big book on transition metals~\cite{KhomskiiBook}, for me always the first thing to do
when encountering a
new problem is to look at what John wrote about it.

\section{Orbital structure of transition metals and its role in exchange interaction}
The first thing to consider when dealing with real
transition metal (TM) compounds is to take into
account more carefully the orbital structure of the respective TM ions. The 5-fold
degenerate $d$-states are split in a crystal by the cubic crystal field
(CF) into the lower three-fold degenerate $t_{2g}$ and the higher-lying two-fold degenerate $e_g$ levels
(for simplicity we will mostly speak about the most
typical local coordination, a TM ion in a ligand (e.g.\ oxygen)
octahedra).
This splitting, often denoted~$10\,Dq$, is typically $\sim1.5$--$2\,\rm eV$
for $3d$ ions, and $\sim2$--$3\,\rm eV$ for $4d$ and $5d$ ones.
Further reduction of symmetry, for example by tetragonal or
trigonal distortions, can lead to further splitting of these levels.

Crucial for the magnetic properties is first of all the type of this
crystal field splitting, and the occupation of particular states by
electrons. Here there are in general two situations possible:
the electrons can occupy the respective levels, starting from those with the
lower energy, following the (first) Hund's rule, i.e.\ first forming
the states with the maximum possible total spin --- the so-called
high-spin states.  Alternatively one can occupy as much as possible the lowest
CF states, e.g.\ the $t_{2g}$ levels, at the expense of ``sacrificing'' the
Hund's rule, which gives the low-spin states. Thus for example ions
with 5 $d$-electrons may fill all CF levels by electrons with parallel
spins, giving the total spin $S=\frac52$ (the high-spin state), or they can fill
three lowest $t_{2g}$ levels, giving the total spin $S=\frac12$ (the low-spin state).
The first situation is most common for $3d$ compounds --- although
here there are some exceptions as well. The second one, the predominant
existence of the low-spin states, is very typical for $4d$ and $5d$
compounds.

Particular orbital occupation, together with the specific geometry,
plays a crucial role in determining the sign and the strength of the
exchange interaction. Corresponding rules were first formulated by
Goodenough in the 50s~\cite{Goodenough1,Goodenough1-2}, and, together with the treatment by
Kanamori~\cite{Kanamori,Kanamori-2,Kanamori-3}, they provide the basis of the modern understanding of
magnetic ordering in different TM compounds. Each time, when
considering new material, the first thing to do is to try to
understand possible types of exchange interactions using the
Goodenough--Kanamori rules.

In the simplest form these rules are the following: if the effective
electron hopping occurs between the half-filled orbitals, with one
electron on each (very often one speaks in this case simply about
filled orbitals, as compared with the empty ones; we will also do it
in what follows), then the situation is similar to that considered in
the Introduction in Eqs.~(\ref{eq:1}) and~(\ref{eq:2}) and illustrated in
Fig.~\ref{FIG:1}(a):
the virtual hopping is allowed only for antiparallel spins at the respective
centres, and the resulting exchange would be antiferromagnetic and
rather strong, $\sim t^2/U$. If however the orbital occupation is such
that the occupied orbital at one centre has nonzero overlap and
hopping only with the empty orbital at the neighbouring site,
Fig.~\ref{FIG:1}(b),
then at first glance it seems that the hopping is allowed
irrespective of the spin orientations. Indeed, as is shown in
Fig.~\ref{FIG:1}(b),
for ``diagonal'' hopping, $t_{11}$ and $t_{22}$ nonzero but the
nondiagonal hopping $t_{12}$ zero, one can always move the electron from
site~$i$ to site~$j$ and back, gaining some energy. But here the
other factor not included in the simple nondegenerate model~(\ref{eq:1}) comes
into play: the Hund's rule exchange interaction, which can be written
as
\begin{equation}
H_{\rm Hund} = -J_{\rm H}\Bigl({\textstyle\frac34} + S_{i,\alpha} S_{i,\beta}\Bigr)\,,
\end{equation}
where $i$ is the site index, and $\alpha$, $\beta$ are the indices of
orbitals ($\alpha \neq \beta$). In this case the energy of the
intermediate state, when we put two electrons e.g.\ on site~$i$ with
parallel spins, $S_{i,\alpha} S_{i,\beta} = \frac14$, is $U-J_{\rm H}$, and that
for antiparallel spins with $S_{i,\alpha} S_{i,\beta} = -\frac34$ is
just~$U$.\footnote{In a more detailed  description one also has to take
into account that the Hubbard interaction~$U$ is different for two
electrons on the same orbital, $U_{11}=U$, and on different orbitals,
$U_{12} = U'<U$. In the spherically-symmetric case $U_{12} = U' = U-2J_{\rm H}$ ---
the so-called Kanamori parametrization. For simplicity we ignore
this factor in what follows; taking it into account usually leads just to
some numerical corrections.} Then the energy gain due to virtual hoppings from a site to
an empty
orbital of a neighbouring site is $-t^2/U$ for antiparallel spins, but
it is larger, $-t^2/(U-J_{\rm H})$, for parallel spins, see Fig.~\ref{FIG:1}(b). We see that in
this case (the hopping between occupied and empty orbitals at
neighbouring centres) the ferromagnetic spin ordering would be
favoured. The difference of these energies determines the strengths
of the resulting ferromagnetic coupling $J\vec S_i\cdot \vec S_j$: for
the usual situation with $J_{\rm H}<U$ (typically for $3d$ elements $U \sim 4\hbox{--}5\,\rm eV$
and $J_{\rm H} \sim 0.8\hbox{--}0.9\,\rm eV$) we will get, keeping the dominant terms, the
resulting ferromagnetic exchange $J \sim -t^2J_{\rm H}/U^2$. That is, in this case
we get ferromagnetic exchange, but it is weaker, reduced by the small
factor $J_{\rm H}/U$ which for 3d ions is ${\sim}\,\frac15$  (but which might be larger
for several unpaired electrons in a shell).

These are, in a nutshell, the two main GK rules: for overlap and hopping
between (half-)filled orbitals we get strong AF exchange, and for
hopping from occupied to empty orbital we get a weaker ferromagnetic
exchange.  We stress that these rules apply to insulators: in
metallic systems there are other mechanisms in play which very often
give a ferromagnetic exchange  (e.g.\ due to the Stoner mechanism~\cite{Stoner}
or due to double exchange~\cite{Zener,deGennes}).

\begin{figure}
\centering
\includegraphics{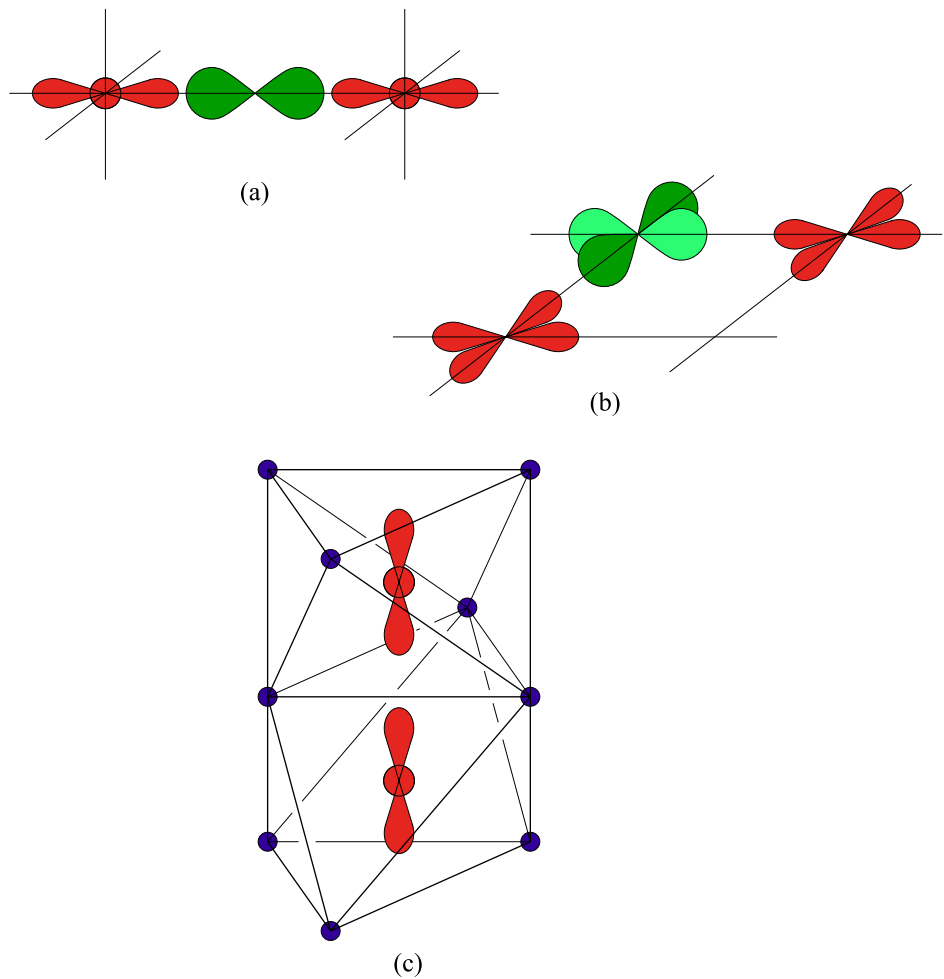}
\caption{\footnotesize\label{FIG:2}Superexchange interaction via $p$-orbitals
of ligands (e.g.\ oxygen). Red are transition metal ions and orbitals,
green are oxygens and their $p$-orbitals.
(a)~$180^\circ$ metal--oxygen--metal ($M$--O--$M$) bonds, as e.g.\ in perovskites
with corner-sharing $M$O$_6$ octahedra. In this case we get rather strong
antiferromagnetic exchange.
(b)~$90^\circ$ $M$--O--$M$ bonds, as e.g.\ for edge-sharing $M$O$_6$ octahedra.
In this case the $d$-orbitals of transition metals overlap with different $p$-orbitals
of oxygen, and the Hund's rule interaction on oxygen gives relatively weak
ferromagnetic exchange. In this case, as well as in case~(c), the direct $dd$-overlap
and hopping is also possible for some orbitals, which can give antiferromagnetic exchange.
(c)~Face-sharing octahedra, with three common oxygens. In this case there exists
superexchange via oxygen $p$-orbitals, but also often rather strong direct $dd$-overlap
and hopping of $a_{1q}$-orbitals, shown in this figure. This direct $dd$-hopping can
give antiferromagnetic exchange, or can lead to metal--metal bonding, see Sec.~4.}
\end{figure}

There are several extra complications in these rules. First of all
most often in the real TM compounds such as oxides the effective
hopping between the $d$-orbitals occurs not directly but via $p$-orbitals of the
intermediate ligands, e.g.\ oxygens. Then, the situation strongly
depends on the local geometry. In the simplest form presented above
these rules work for the direct $dd$-hopping or, more often, for
the neighbouring $M$O$_6$ octahedra with the common corner, one common
oxygen, with the hopping via oxygens,s in the case of $180^\circ$
metal--oxygen--metal bonds, as e.g.\ in perovskites,
Fig.~\ref{FIG:2}(a).
In this case one often describes the resulting situation as the
rule ``same orbitals --- opposite spins; different orbitals --- same
spins'', or, expressed differently, ``ferro orbitals --- antiferro
spins; antiferro orbitals --- ferro spins''. But one has to be very
careful in using this ``rule of thumb'': it is indeed valid for the
geometry with the $180^\circ$ $M$--O--$M$ bonds, but it is not valid at all
for example for another rather common situation, with the octahedra
with a common edge and with $\sim90^\circ$ $M$--O--$M$ bonds,
Fig.~\ref{FIG:2}(b).
All these rules are described in details e.g.\ in~\cite{GoodenoughBook,KhomskiiBook,Streltsov2}.

Strangely enough, the third typical situation --- with the neighbouring
$M$O$_6$ octahedra with a common face, with three common ligands (oxygens),
Fig.~\ref{FIG:2}(c)
--- is very rarely considered in the literature; see for
example the recent treatment of all three situations, with the
emphasis on the common face geometry, in~\cite{Kugel}. Interestingly enough,
the situation with the common face usually resembles more that with
the common corner than the one with the common edge.

All in all, using these general rules, largely formulated by
Goodenough, extended to
include the details of the crystal structure and local geometry,
and including when necessary the $p$-orbitals of ligands in an apparent way
(see Sec.~6 below), turns out to be extremely successful in
explaining the main features of magnetic exchange, and consequently many
other properties, of an enormous amount of magnetic materials. Indeed
the development which followed the classical works of Goodenough,
Kanamori and Anderson, not only laid the groundwork for the huge
field of magnetic insulators, but for many years were very
successfully applied, and will be certainly applied in future, to
explain properties of many magnetic systems, both known and those yet to
be discovered. Thus one can really say that this development is a
glorious chapter in the history of magnetism.

\section{Jahn--Teller effect and orbital ordering}
A very special situation which is often encountered in the physics of TM
compounds is that with orbital degeneracy. When taking the orbital
structure into account and filling the respective energy levels by the
required number of electrons, we can end up in a situation in which
an electron occupies one of several degenerate orbitals, e.g.\ when we have
one electron (or one hole) in doubly-degenerate levels. The first
situation (one electron) is met e.g.\ for Mn$^{3+}$ in a regular
octahedron, with the configuration $t_{2g}^3e_g^1$; the second one, with one
hole, for Cu$^{2+}$ ($t_{2g}^6e_g^3$).  It was shown by Jahn and Teller~\cite{JahnTeller}
that this situation is absolutely unstable: in this case the system
spontaneously develops distortions reducing the symmetry and leading to a splitting of
the degenerate levels.  This is the famous Jahn--Teller (JT) effect.\footnote{Interestingly
enough, as is described by Teller himself, the idea
of this effect was suggested to him by Landau, see e.g.\ Appendix~A.2
in~\cite{KhomskiiBook}.}
(I formulate here the essence of the JT effect in a very simplified and
not very rigorous form, but it is sufficient for our purposes; see a
more detailed discussion in the very big literature on JT effect,
e.g.\ in~\cite{Bersuker}.) This distortion and the resulting splitting
of degenerate orbitals occurs in concentrated systems as a real
structural phase transition, which we can call the cooperative JT
effect (for local JT centres the situation is much more complicated~\cite{Bersuker}.).
It leads to the occupation by
electrons of particular orbitals, e.g.\ one of two possible $e_g$
orbitals, i.e.\ one can say that in this case there occurs in a system
an orbital ordering. Nowadays in the physical literature we indeed
more often speak about orbital ordering (O.O.) than about the
cooperative JT effect, although one should stress that these are just
two different sides of the same phenomenon, the one (orbital
ordering) cannot exist without the other (JT distortions).

\begin{figure}
\centering
\includegraphics{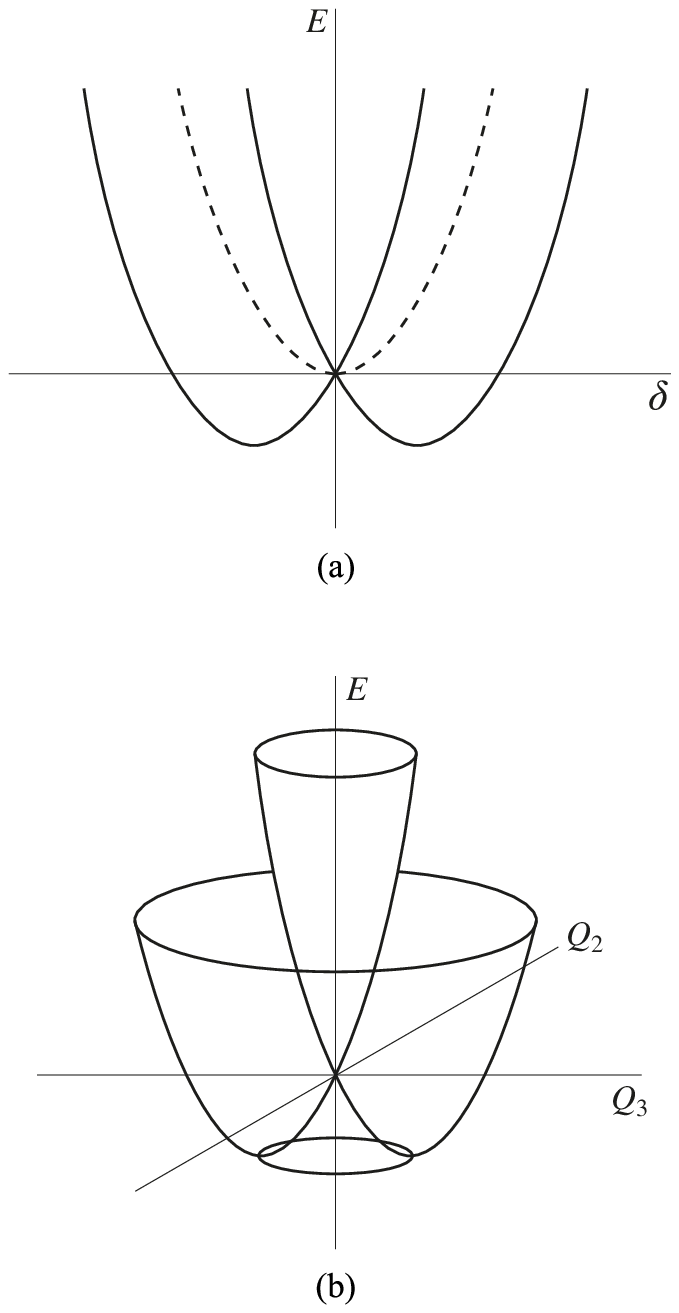}
\caption{\label{FIG:3}\footnotesize Jahn--Teller effect for double orbital degeneracy, e.g.\
for $e_g$-electrons.
(a)~Schematic explanation of the Jahn--Teller effect.
(b)~The form of the energy surface (``Mexican hat'') for double-degenerate $e_g$-electrons
interacting with $E_g$ distortions (tetragonal $Q_3$ mode, orthorhombic $Q_2$ mode).
Note the continuous degeneracy of the lowest energy states (the trough of the Mexican hat)
and the presence of the conical intersection, both leading to important quantum effects.}
\end{figure}

The idea of the JT effect can be illustrated by the simple treatment,
Fig.~\ref{FIG:3}:
The energy of two states, degenerate for the undistorted system,
as a function of distortion~$\delta$ has the form
\begin{equation}
E = -g\delta + \sfrac12B\delta^2\,.
\end{equation}
Here the first term is the splitting of degenerate electronic levels
by the distortion (reminiscent of the Zeeman splitting in a magnetic
field), and the second term is the elastic energy of the lattice.
One immediately sees that the symmetric, undistorted situation with
degenerate levels is always unstable, it is never the minimum of the
energy. Minimization of this expression gives for the equilibrium
distortion the value $\delta_0=-g^2/B$, and the energy at the minimum
is $E_{\rm JT} = - g^2/2B$. In concentrated systems the distortions at
different centres interact with each other (e.g.\ by elastic forces),
which leads to the cooperative effect: distortions of different
centres occurs simultaneously, leading to a real phase transition ---
which is simultaneously a structural transition and an orbital
ordering. The mechanism of this phenomenon described above is the JT
electron--lattice interaction.

In real situations usually the degenerate electron states can be
split not by one but by several types of distortions. Thus the
doubly-degenerate $e_g$ levels can be split by tetragonal distortions
$Q_3$, but also by orthorhombic distortions $Q_2$ (distortions of the $E_g$
symmetry). In effect the energy surface will be the one shown in
Fig.~\ref{FIG:3}(b)
(it can be visualised as the surface obtained by the rotation of
the energy of Fig.~\ref{FIG:3}(a) about the vertical axis). This is the famous
``Mexican hat'' appearing in many field of physics, including the Higgs
phenomenon. This situation leads to many nontrivial quantum effects
connected with orbitals, see e.g.~\cite{Khomskii2,Bersuker} and Sec.~9 below.

Interestingly enough, there exists also another, purely electronic
mechanism of such transition, closely related to the physics
described in the previous section. Indeed, suppose we have two
neighbouring centers with doubly-degenerate orbitals but with one
electron per site. Suppose that the situation with orbital overlap
and intersite electron hopping is like the one shown in Fig.~\ref{FIG:1}(b): the electrons
can hop from one orbital only to the same orbital at neighbouring
sites, $t_{11}=t_{22}=t$, $t_{12}=0$. In this case, instead of two possible
situations shown in Fig.~\ref{FIG:1}(a), leading to the antiferromagnetic
superexchange, we have four possibilities: same orbital--same spin;
same orbital--different spins; different orbitals--same spin;
different orbitals--different spins, see Fig.~\ref{FIG:1}(b). The arguments
of the previous section show that the biggest energy gain
due to virtual hopping of electrons between sites is reached in this
case for the third case shown in Fig.~\ref{FIG:1}(b), i.e.\ different orbitals--same spins.
That is, the same superexchange process which for the
nondegenerate case gave the effective antiferromagnetic exchange~(\ref{eq:2})
and the corresponding spin ordering, in this case would give
simultaneously both spin and orbital orderings, in this particular
situation ferro spin and antiferro orbitals. This is just the ``rule of
thumb'' mentioned in the previous section. (We stress once again that
it is valid for this particular model, for certain particular
geometries, etc.)   Mathematically one would then describe the coupled
spin and orbital exchange by an effective Hamiltonian of the type
\begin{equation}
H = J_1 \sum_{ij} \vec S_i\cdot\vec S_j  + J_2 \sum_{ij} \tau_i\tau_j + J_3 \sum_{ij} (\vec S_i\cdot \vec S_j)(\tau_i\tau_j)\,,
\label{eq:5}
\end{equation}
where the $\tau$ operators (operators of the pseudospin~$\frac12$) describe doubly-degenerate orbitals
(here spins enter as a scalar product $\vec S_i\cdot\vec S_j$, but the structure of orbital terms may
be more complicated~\cite{KugelKhomskii}).

This purely electronic, or exchange mechanism of coupled spin and
orbital ordering was first proposed in~\cite{KugelKhomskii2,KugelKhomskii2-2,KugelKhomskii}, and
it is sometimes called the Kugel--Khomskii model. It seems that it can
indeed give purely orbital ordering without the lattice involved. But
we stress that as soon as we include coupling with the lattice ---
which of course is always present in real situations --- the lattice
would immediately follow, so that the structural transitions of this
type and orbital ordering always occurs simultaneously. The
difference is just in the dominant mechanism of this transition: it
may be predominantly of electron--lattice type or of the exchange type. In
real systems both these mechanisms of course act simultaneously, so
that it is difficult to tell which one is more important. One can
get this information from {\it ab~initio} calculations in which one can
artificially ``switch off'' the JT interaction. The corresponding calculations
show that typically these two mechanisms, electron--lattice (JT)  and
superexchange (Kugel--Khomskii) ones, are of comparable magnitude, the JT
mechanism usually being somewhat stronger (with the ratio of about
60:40 or 70:30)~\cite{Pavarini,Pavarini-2}.

Orbital ordering has many manifestations, and it can lead to many quite
nontrivial effects which are widely studied experimentally and
theoretically, see e.g.~\cite{Tokura,Khomskii2}.

\begin{figure}
\centering
\includegraphics{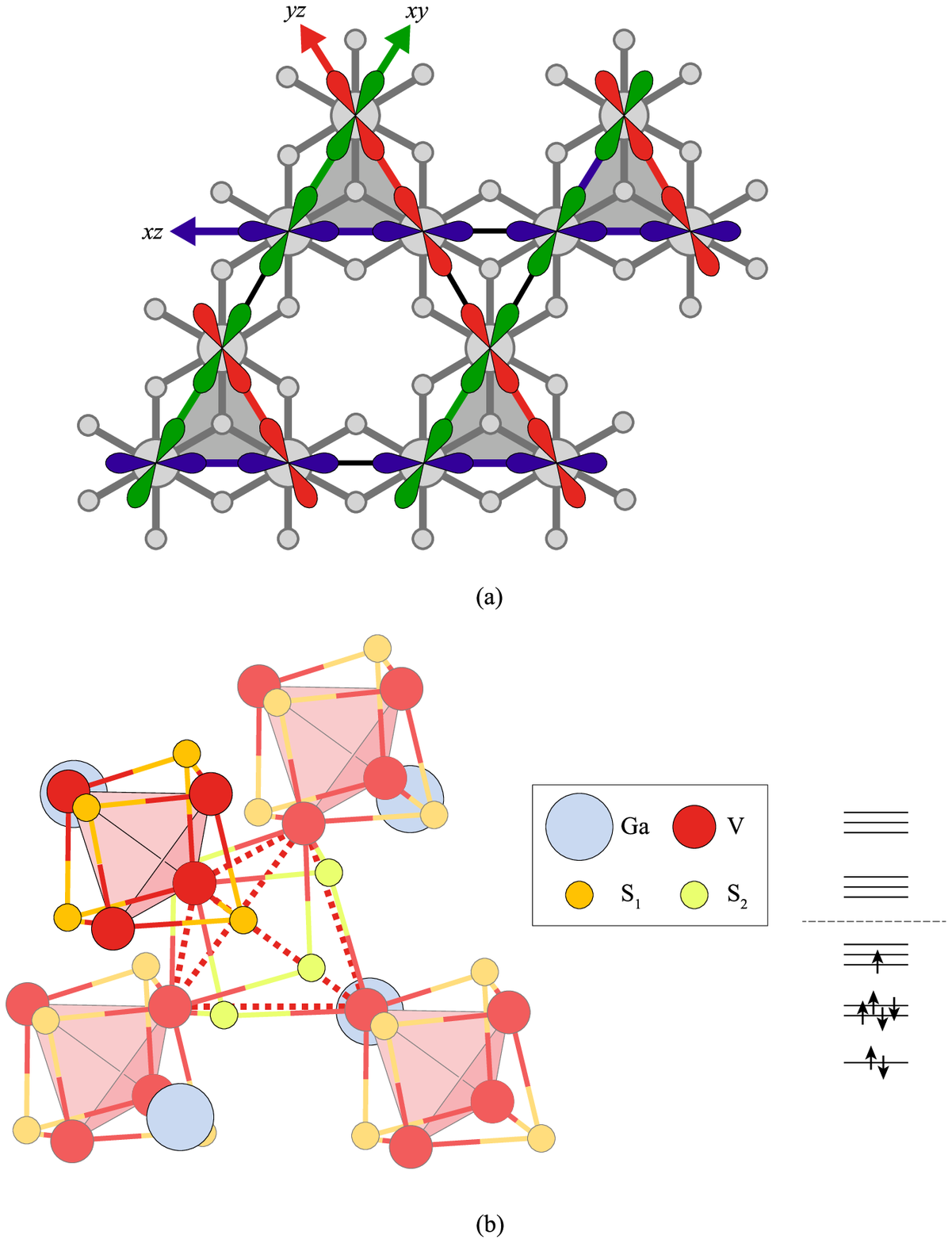}
\end{figure}

\begin{figure}
\centering
\includegraphics{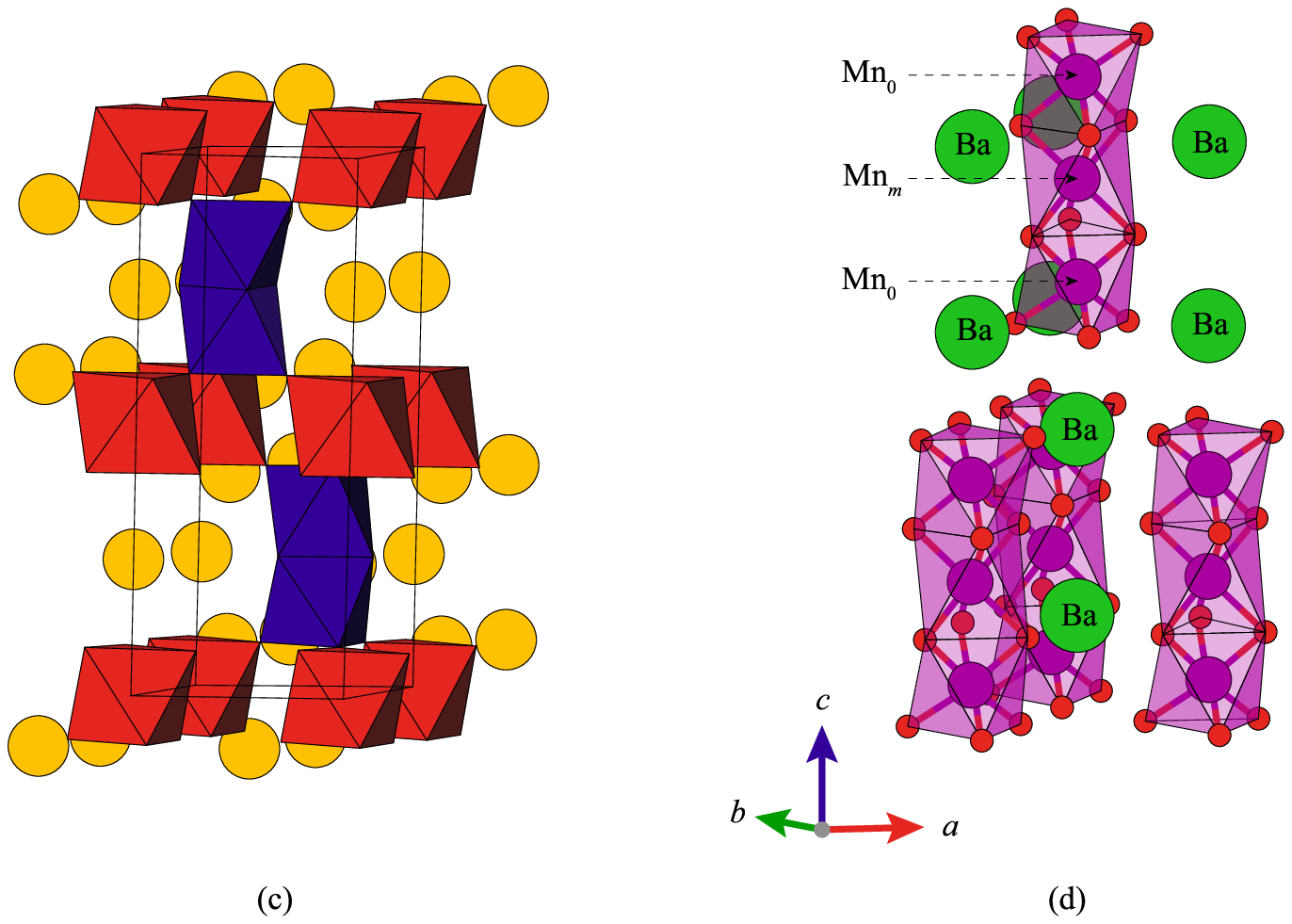}
\caption{\label{FIG:4}\footnotesize Several typical cases of systems with different types
of crystal lattices and with molecular-type clusters of transition metals.
(a)~Breathing kagome lattice as e.g. in Zn$_2$Mo$_3$O$_8$~\cite{Wang,Wang-2,Wang-3,Wang-4}
or Na$_2$Ti$_3$Cl$_8$~\cite{Kelly}, with alternating small (shaded) and large triangles.
(b)~Breathing pyrochlore lattice, as in a lacunar spinel CaV$_4$S$_8$. Pink are small tetrahedra. On the right
is the energy diagram of this system. For 7 electrons per small V$_4$S$_4$ cluster
there exists an orbital degeneracy (one electron in the higher-occupied triplet),
and the resulting Jahn--Teller distortion makes GaV$_4$S$_8$ ferroelectric and multiferroic~\cite{Ruff,Ruff-2}.
(c)~Materials of the class Ba$_3MM'_2$O$_9$, with $M = \rm Na$, Ca,~Ti, occupying single
$M$O$_6$ octahedra (red) and with transition metal $M'$ ($=$~Ru, Ir,~\dots) forming
dimers (blue) with common edge of the $M'$O$_6$ octahedra, see e.g.~\cite{Kimber,Kimber-2}.
(d)~Materials of the class Ba$_4$Nb$M_3$O$_{12}$ ($M = \rm Mn$, Ru, Rh,~Ir), containing
trimers made of face-sharing $M$O$_6$ octahedra~\cite{Nguyen,Komleva}.
}
\end{figure}

\section{``Molecules in solids'': orbital Peierls transitions}
A very special class of systems and effects, in which orbital degrees
of freedom play an important role, is provided by materials with
particular tightly-bound metal clusters. Such clusters may be dimers,
trimers, or larger clusters. They may exist just because of a
specific crystal structure of a material, or may be formed
spontaneously below some phase transition. The study of these
materials became quite active very recently. There are many examples
of such systems nowadays. These are for example the so-called
``breathing'' lattices --- breathing kagome-like
(Zn,Fe,Ni)$_2$Mo$_3$O$_8$~\cite{Wang,Wang-2,Wang-3,Wang-4} or Na$_2$Ti$_3$Cl$_8$~\cite{Kelly}, with
alternating small and large triangles, Fig.~\ref{FIG:4}(a); breathing
pyrochlores such as GaV$_4$S$_8$~\cite{Ruff,Ruff-2} with ordered large and small
tetrahedra, Fig.~\ref{FIG:4}(b). There exist systems with well-separated dimers
--- Ba$_3M$Ru$_2$O$_9$ or Ba$_3M$Ir$_2$O$_9$ ($M=\rm Na$, Ca, In, Ti,~\dots)~\cite{Kimber,Kimber-2}, Fig.~\ref{FIG:4}(c);
trimers --- Ba$_4$Nb$M_3$O$_{12}$  ($M = \rm Mn$, Rh, Ir)~\cite{Nguyen,Komleva},
Fig.~\ref{FIG:4}(d); etc. In some cases tightly bound dimers are formed below
phase transitions, as e.g.\ in VO$_2$~\cite{Morin,Goodenough2}, MgTi$_2$O$_4$~\cite{Schmidt},
CuIr$_2$S$_4$~\cite{Radaelli}, or trimers formed below $T_c$, e.g.\ in LiVO$_2$~\cite{Kobayashi,Pen},
LiVS$_2$~\cite{Katayama}.

Detailed examination of most of such examples shows that the orbital
structure of TM ions forming such clusters, ``molecules in solids'',
plays here a very important role. Typically such clusters are formed by
metal--metal bonds, created by direct $dd$-overlap in materials with $ML_6$
octahedra with a common edge or common face, in which the TM ions are
relatively close to each other. Often in such cases the metal--metal
distance in clusters is indeed quite small --- comparable
or even shorter than in a respective metal. Thus the short Ru--Ru
distance in Ru dimers in Li$_2$RuO$_3$ is $2.568\,$\AA~\cite{Miura}, whereas
such distance in Ru metal is $2.65\,$\AA; and orbital ordering plays a very
important role in this system~\cite{Jackeli}. The V--V distance in V trimers
(triangles) in LiVO$_2$ is $2.56\,$\AA~\cite{Kobayashi}, and that in V metal it is $2.62\,$\AA
(note that this short distance in LiVO$_2$ is also much smaller than
the critical value separating localised and itinerant states, which,
according to Goodenough~\cite{GoodenoughBook}, is~$2.94\,$\AA).  And in these situations
the $d$-orbitals pointing to the neighbouring TM ions, cf.\ Fig.~\ref{FIG:2}(a), can indeed form
strong bonding (and antibonding) states, i.e.\ form molecular orbitals
as in real molecules, and it is this bonding that stabilises the very
cluster --- Ru dimers in Li$_2$RuO$_3$, Ir dimers in CuIr$_2$S$_4$, V dimers in VO$_2$
and MgV$_2$O$_4$, trimers in LiVO$_2$ or Zn$_2$Mo$_3$O$_8$, etc. Such ``chemical'' description is usually very
illuminating and helps to understand the origin of such clusters and
explain their electronic structure and properties. Thus for example
the electronic structure of tightly-bound V tetrahedra in the lacunar
spinel GaV$_4$S$_8$ (which can be written as Ga$_{1/2}$(Vacancy)$_{1/2}$V$_2$S$_4$ --- a
spinel with ordered Ga ions and vacancies at $A$-sites) can be well
explained by the picture of molecular orbitals, shown in Fig.~\ref{FIG:4}(b):  7~electrons in this
cluster fill these levels as shown in this figure, and this level
filling explains why the net spin of such cluster is $S=\frac12$. Moreover,
we also see that the last, seventh electron here occupies one of
the triply-degenerate levels, which, by the physics explained in the previous
section, can lead to the JT effect. And indeed there exists a structural
transition in GaV$_4$S$_8$, which lifts this degeneracy, and below
this transition this material becomes ferroelectric, i.e.\ actually
multiferroic~\cite{Ruff,Ruff-2}.

\begin{figure}[ht]
\centering
\includegraphics{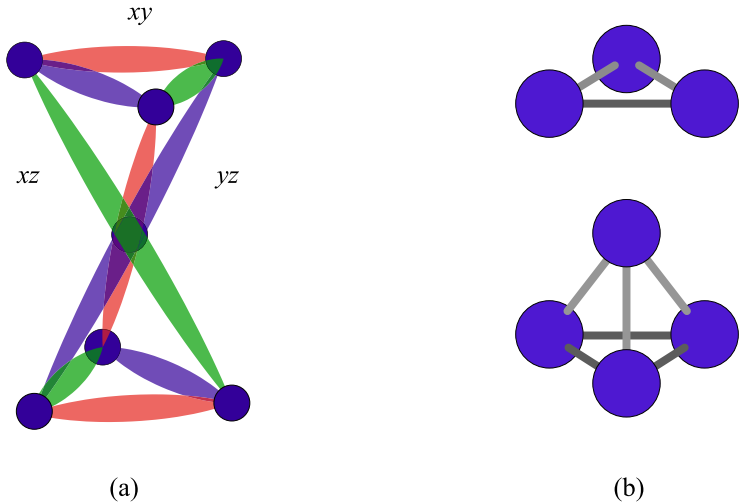}
\caption{\label{FIG:5}\footnotesize Superstructure in AlV$_2$O$_4$ below the structural
transition. Initial interpretation was in terms of heptamers --- clusters of 7 Vanadium
ions,~(a)~\cite{Horibe}. Recent local measurements demonstrated that the middle V ion shifts
towards one of the base triangles, forming V tetrahedra and triangles,~(b)~\cite{Browne}. In both
pictures the particular orbital occupation plays a crucial role.}
\end{figure}

The ``molecular'' picture can also explain the formation of a very
exotic structure of the spinel AlV$_2$O$_4$, with mixed valence of~V.
Initially it was claimed~\cite{Horibe} that below the phase transition
there are V heptamers formed --- clusters made of 7 V ions,
Fig.~\ref{FIG:5}(a). But more detailed local measurements (using pair
distribution function analysis) have shown that in fact there are not
heptamers but rather trimers --- V triangles and V$_4$ tetrahedra,
Fig.~\ref{FIG:5}(b)~\cite{Browne}. And this can be naturally explained in the
picture in which V$^{3+}$ ions with two $d$-electrons form triangles, with
two valence bonds per each~V, formed by using the corresponding V
orbitals; and V$_4$ tetrahedra are made of V$^{2+}(d^3)$ ions, with each
electron of V$^{2+}$ forming its own chemical bond, three bonds per each
V in a tetrahedra (V ions in AlV$_2$O$_3$ have mixed valence~$V^{2.5+}$).

\begin{figure}[ht]
\centering
\includegraphics{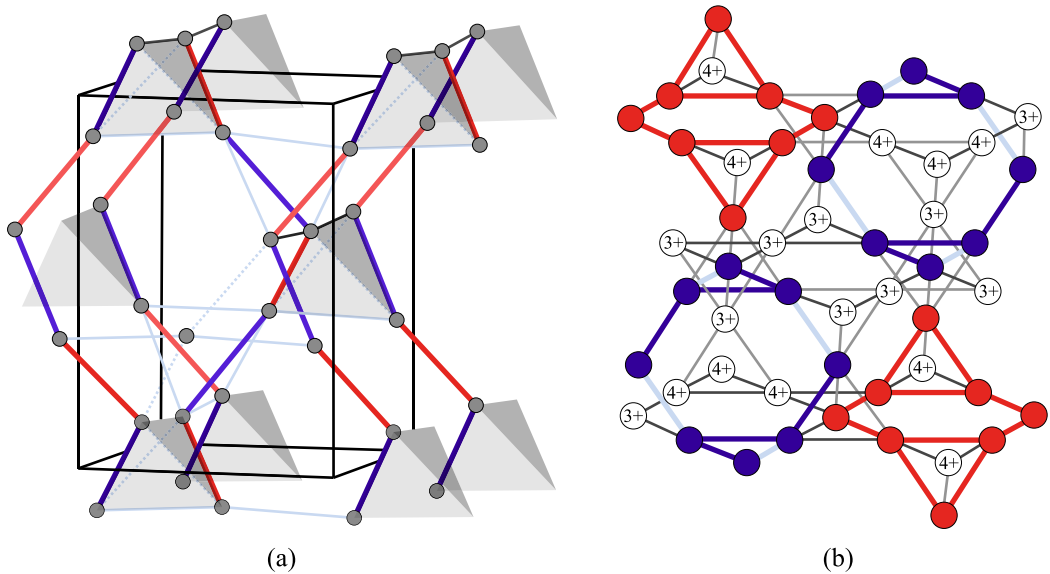}
\caption{\label{FIG:6}\footnotesize Crystal structure of MgTi$_{2}$O$_4$~(a)~\cite{Schmidt} and CuIr$_2$S$_4$~(b)~\cite{Radaelli}
below phase transitions. Short Ti--Ti and Ir--Ir dimers are formed in the low-temperature phase,
with the use of orbitals directed from one TM to the next.}
\end{figure}

\begin{figure}[ht]
\centering
\includegraphics{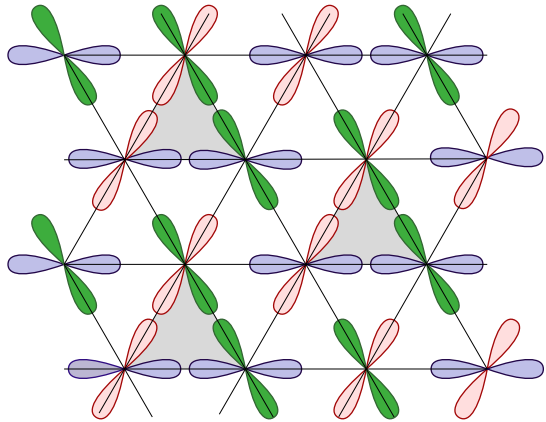}
\caption{\label{FIG:7}\footnotesize Trimerization in LiVO$_2$~\cite{Kobayashi}. It
can be explained either in a ``chemical'' picture, as the formation of strong
valence bonds in small V$_3$ triangles (shaded)~\cite{Pen}, or as a result of Peierls trimerization in $\frac13$-filled
one-dimensional bands~\cite{KhomskiiComment,KhomskiiComment-2}.}
\end{figure}

Interestingly enough, often one can also describe the origin of such
clusters in a different picture --- the picture of the orbital Peierls
state~\cite{Mizokawa,KhomskiiComment,KhomskiiComment-2}. Peierls instability exists
in the original form in one-dimensional systems and leads to a
distortion of a regular chain such that it opens a gap at the
corresponding Fermi-surface. Thus e.g.\ for one electron per site the
tight-binding band is half-filled, and dimerization of such chain
would open a gap at the Fermi-surface (for 1d case --- the Fermi-points).
One can visualise this process as a first step to the
formation of a molecular solid: e.g.\ if this chain with one electron
per site is a chain of hydrogen atoms, the H$_2$ dimers formed at such
Peierls dimerization are the first step in the formation of H$_2$
molecules. But similar distortion can also exist for other electron
concentrations and other band fillings: for $\frac13$- or $\frac23$-filled bands
one would get trimerization, and for $\frac14$-filled 1d bands one would get
tetramerization. Apparently one can use this picture to rationalise
the formation of many superstructures and molecular clusters in many
systems. Thus, for example, due to specific geometry different
$d$-orbitals form quasi-one-dimensional bands in spinels such as MgTi$_2$O$_4$,
and $\frac14$-filling of these bands leads to tetramerization in
the corresponding directions, which naturally explains~\cite{Mizokawa}
beautiful superstructures observed in MgT$_2$O$_4$~\cite{Schmidt} and in CuIr$_2$S$_4$~\cite{Radaelli},
Fig.~\ref{FIG:6}. Similarly, one can explain the formation of trimers in
breathing kagome systems (Zn,Fe,Ni)$_2$Mo$_3$O$_8$ and
Na$_2$Ti$_3$Cl$_8$ by dimerization in half-filled bands in 1d chains
in kagome lattice, Fig.~\ref{FIG:4}(a) (blue, red and green bands there) and the formation of trimers in LiVO$_2$ and
LiVS$_2$ with triangular lattices --- as a consequence of trimerization
of $\frac13$-filled 1d bands~\cite{KhomskiiComment,KhomskiiComment-2}, Fig.~\ref{FIG:7}.  We see that the
orbital degrees of freedom can be responsible for the formation of
very exotic superstructures --- ``molecules in solids'', observed in
quite a few TM compounds, which have many nontrivial and potentially
useful properties (such as e.g.\ multiferroicity and the formation of
skyrmions in lacunar spinels~\cite{Ruff,Ruff-2}).  This direction of research,
which is rapidly developing right now, is thus one more example of
the rich physics related to orbitals on TM compounds.

\section{Orbital-selective effects}
The directional character of orbitals, which was very important for
the GKA rules considered in Sec.~2 and which gave directional ``chemical
bonds'' in clusters described in Sec.~4, can also lead to other
nontrivial effects. Thus, it may happen that the effective overlap
and the corresponding hopping of electrons on different orbitals can be
of very different magnitude: some orbitals can overlap very strongly,
whereas other electrons on other orbitals can have very weak hopping
to neighbouring sites. As explained in the Introduction, the very
existence of localised vs itinerant states of electrons in solids
crucially depends on the value of this hopping and of the respective
bandwidth: for hopping $t>U$ we have itinerant electrons, i.e.\ the metallic
state, whereas for $t<U$ we have localised electrons, Mott insulators. And in
the situation in which electrons in different orbitals have very
different hoppings~$t$, it may so happen that for one of these
orbitals we have $t>U$, but for the others~$t<U$. In this case we can
expect different states for these different electrons, i.e.\ the
itinerant and localised electrons  (all of these are $d$-electrons!)\ could
coexist in one system. This was the idea of orbital-selective Mott
transition, first put forth in~\cite{Anisimov} and later
extensively discussed in the literature, see e.g.~\cite{Georges}.

\begin{figure}
\centering
\includegraphics{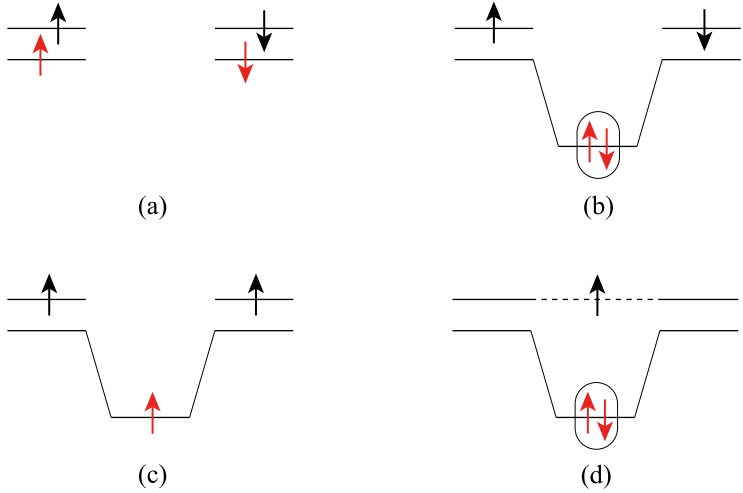}
\caption{\label{FIG:8}\footnotesize Orbital-selective metal--metal bonding, by~\cite{Streltsov3}.
(a),~(b)~The situation with two electrons per site on two orbitals, one of which has a strong overlap and
intersite hopping~$t$, whereas the other has small or zero hopping.
In (a) the situation with the strong Hund's rule coupling is shown, $J_{\rm H} > t$,
which leads to the formation of $S=1$ states at each centre, with some
antiferromagnetic interaction between those.
(b)~The situation with strong hopping $t>J_{\rm H}$, leading to the formation
of a singlet bond made of strongly overlapping orbitals, with the remaining
localized electrons with $S=\frac12$ per site.
(c),~(d)~Similar situation in case of three electrons per dimer.
(c)~The situation with strong Hund's exchange, $J_{\rm H}> t$, leading
to parallel orientation of all three spins: two localized one per site, and one ``itinerant'',
hopping from site to site. This is the double exchange mechanism of ferromagnetism
e.g.\ in metallic oxides such as colossal magnetoresistance manganites.
(d)~The situation with very strong intersite hopping on ``itinerant'' orbitals, $t>J_{\rm H}$.
In this case two electrons form a singlet bonding state made of these orbitals,
and there remains one localized electron per dimer, with $S=\frac12$ (in contrast
to the total spin of a dimer $S=\frac32$ in the case~(c)). Thus in this situation
the formation of orbital-selective molecular orbital on a dimer leads to the suppression of double exchange.}
\end{figure}

The same effect may also exist and may even have stronger
manifestations in finite clusters. This was studied in~\cite{Streltsov3},
where it was shown that the same physics can lead to a strong
suppression of magnetic moment of TM ions and can result in strong
violation of the Hund's rule. Indeed, if we consider a pair of sites,
a dimer, such that one orbital at each site has strong overlap with
the neighbour but the other(s) have much smaller overlap, or no overlap at all,
then for example for two orbitals and two electrons per site we can have two possible
situations, Fig.~\ref{FIG:8}. If the on-site Hund's interaction is large enough,
these two electrons at each site would each form a spin-1 state, and
intersite hybridization would not destroy such $S=1$ states; there
would be a Heitler--London-like state formed by these $S=1$ states, Fig.~\ref{FIG:8}(a).
However if the intersite hopping~$t$ of one orbital, e.g.\ with
the lobes directed towards the neighbouring site, is large enough, we
can form a different state: the electrons on the ``active'' orbital would
form a bonding state with energy~$-t$, and two electrons would
occupy this bonding state forming a singlet on it, Fig.~\ref{FIG:8}(b). And the
remaining one electron per site would remain localised, but the
moment of the site would now be not $S=1$ but strongly reduced, $S=\frac12$.
This could happen if the bonding energy $\sim t$ exceeds the Hund's rule
intraatomic exchange $J_{\rm H}$: the strong bonding of one orbital leads to
a violation of the first Hund's rule. This is hardly possible for $3d$
systems but can easily happen in $4d$ and $5d$ materials, for which
the electron hopping~$t$ increases but the Hund's coupling~$J_H$ decreases.
Such orbital-selective formation of bonding states (or molecular
orbitals), with strong suppression of the moment per site, was
observed in many materials containing dimers. Thus for example in
Y$_5$Re$_2$O$_{12}$~\cite{Chi},
with three electron per site, the moment per site is not $\frac32$ but only
$\frac12$: apparently two orbitals form a bonding singlet states, and only
one ``free'' localised spin remains at each site.

This effect may be even more pronounced for partial
occupation of sites, in the situation which usually leads to double
exchange (DE) and to net ferromagnetism. The simplest model of DE is
the model of two sites with three electrons; actually it was on this
model that Zener first proposed the idea of DE~\cite{Zener}.
In this case, as is shown in Fig.~\ref{FIG:8}(c),~(d), there are also two
options: for strong  Hund's coupling (this is the assumption always
made in the treatment of DE!)\ we indeed gain more energy if all spins
are parallel; i.e.\ in this case we have a ferromagnetic state. But if
for one orbital the intersite hopping is large enough, bigger than
$J_{\rm H}$, then we can also first form from these orbitals the bonding
state and put on this state two electrons with opposite spins. In
this case a dimer would have only a free spin $S=\nobreak\frac12$, Fig.~\ref{FIG:8}(d),
instead of $S=\frac32$ as in the DE state of Fig.~\ref{FIG:8}(c). That is, strong hopping
on one orbital can effectively suppress the DE mechanism of
ferromagnetism. Thus the orbital selectivity can have a profound
influence on the magnetic state of a system, effectively suppressing the
values of on-site magnetic moments and even long-range
ferromagnetism.

One can also think of other possible effects connected with very
different behaviour of electrons on different orbitals and with the
directional character of orbitals. It is this directional dependence
which gives rise in particular to orbital-dependent exchange as in
the compass and Kitaev models, see Sec.~7 below; it is yet another
manifestation of specific effects of orbitals in TM compounds.

\section{Charge-transfer gap systems}
As mentioned in Sec.~2, in most interesting magnetic insulators the TM
ions are surrounded by a ligand cage, and the effective hopping of
$d$-electrons, responsible, in particular, for the exchange
interaction, occurs via the intermediate ligands, usually via their
$p$-orbitals, e.g.\ the $2p$-orbitals of oxygens, which we will mainly have in
mind. In this case in principle one should include these $p$-electrons
in the theoretical description, and not confine oneself to
$d$-electrons, as was done in the Hubbard model~(\ref{eq:1}). Instead one
should, generally speaking, use the $d$--$p$ model, which schematically
looks like
\begin{equation}
H =     \sum_{ij,\sigma} \epsilon^{\vphantom{\dagger}}_d d_{i\sigma}^\dagger d^{\vphantom{\dagger}}_{j\sigma}
      + \sum_{ij,\sigma} \epsilon^{\vphantom{\dagger}}_p p^\dagger_{i\sigma} p^{\vphantom{\dagger}}_{j\sigma}
      + \sum_{ij,\sigma} t_{pd} (p^\dagger_{i\sigma} d^{\vphantom{\dagger}}_{j\sigma} + \hbox{h.c.})
      + U\sum_i n_{di\uparrow}n_{di\downarrow}\,.
\label{eq:6}
\end{equation}
In principle one has to include here the orbital indices of
the respective $d$- and $p$-orbitals, and, when necessary, also other
interactions --- e.g.\ the Hund's exchange, Hubbard-like repulsion of
$p$-electrons $U_{pp}$, etc.

\begin{figure}[ht]
\centering
\includegraphics{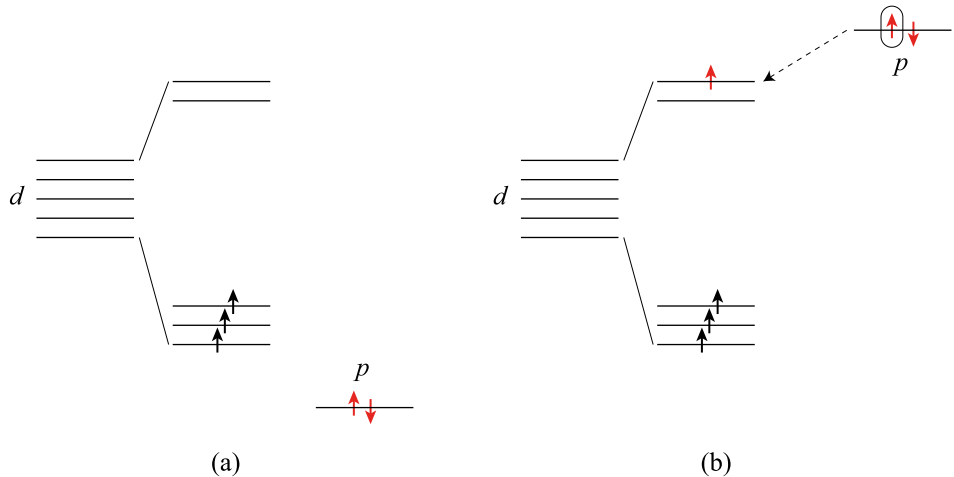}
\caption{\label{FIG:9}\footnotesize Schematic energy structure of transition metal systems
with the apparent inclusion of ligand (e.g.\ oxygen) $p$-holes
(a)~The usual situation with deep $p$-levels and large charge-transfer gap.
(b)~The situation with small or negative charge-transfer gap, in which case there may occur ``self-doping''
--- spontaneous transfer of one of $p$-electrons of oxygen to the $d$-shell of transition metal,
leading to the formation of oxygen holes. Note that this is a simplified picture; in fact the energies
of respective $p$- and $d$-levels should include the contribution of electron--electron interactions,
notably the Hubbard's~$U$, see text.}
\end{figure}

In this situation, which was also considered in great detail by
Goodenough in many papers and in the book~\cite{GoodenoughBook}, an important
parameter, besides the electron hopping and the Hubbard~$U$ (and Hund's
coupling~$J_{\rm H}$), is the energy difference between $d$- and $p$-levels,
i.e.\ the difference between the ``original'' configuration $d^np^6$
(with certain valence of TM and with oxygen O$^{2-}$) and the excited
state in which we transfer an electron from oxygen to TM --- the charge
transfer excitation energy
\begin{equation}
\Delta_{\rm CT} =  E(d^{n+1}p^5) - E(d^np^6)  \,.
\label{eq:7}
\end{equation}
In the description with the Hamiltonian~(\ref{eq:6}) it may look as though
$\Delta_{\rm CT} = \epsilon_d - \epsilon_p$,  but in fact one has to
include in the definition of charge transfer energy also the effects
of the change of interactions, such as on-site Coulomb
interactions on $d$- and $p$-levels, implicitly taken into account in~(\ref{eq:7}).
But it is often possible to use
the schematic one-electron picture with single-particle $d$- and $p$-levels, as in Fig.~\ref{FIG:9}
--- remembering that in fact these are the levels of particular
many-electron configurations.

In this more general and more realistic description one would have in
the exchange processes not only virtual transitions between $d$-levels,
of the type $d^nd^n \leftrightarrow d^{n-1}d^{n+1}$, which cost energy~$U$ (this is the
$U$ in the denominator of the exchange integral $J$ in~(\ref{eq:2})); but there
would be also processes of virtual transitions to and from
intermediate oxygens, $d^np^6 \leftrightarrow d^{n+1}p^5$, with the excitation energy
$\Delta_{\rm CT}$~(\ref{eq:7}). All this is now a well-developed field, and, again,
orbital occupation plays a very important role in determining the
exchange processes and in determining the resulting exchange
constants.

Very special situations may appear if the energies of these
configurations, $d^np^6$ and $d^{n+1}p^5$, becomes comparable, so that the
charge transfer energy $\Delta_{\rm CT}$ is very small or even becomes
negative, Fig.~\ref{FIG:9}(b).  In this case, first of all, it is just this
charge transfer transition which would dominate in exchange
processes.  As a result the exchange integrals as in (\ref{eq:2}) would
contain in the denominators not the Hubbard's~$U$ but the charge
transfer energy $\Delta_{\rm CT}$~\cite{GoodenoughBook,KhomskiiBook}.  But for negative CT energy
there may also occur stronger modifications: there may exist
``self-doping''~\cite{Korotin} --- some electrons may spontaneously go from
oxygens to transition metals, so that there would be oxygen holes
created even without external doping. In this case the material may
even become metallic. It is a very interesting and rich, though often
obscure field, which develops now rather rapidly, see e.g.~\cite{Khomskii3,Sawatzky}.

In such situation with small or negative CT gap the typical
orbital effects such as the JT effect and orbital ordering, may also
change. Thus e.g.\ the detailed form of orbital ordering for the case
of $e_g$ degeneracy may change~\cite{Mostovoy}. And in some
cases the presence of ligand holes can even suppress the otherwise very
strong JT effect, even in such famous JT ions as Cu$^{2+}$, which happens
e.g.\ in Ba$_3$CuSb$_2$O$_9$~\cite{Takubo}. Thus the orbital effects turn out
to be rather sensitive to the details of the electronic structure, especially
in the case of small or negative CT energy.

\section{Orbitals, frustrations, spin and orbital \hbox{liquids}}
The directional character of orbitals, already stressed above, can also
cause very special effects in magnetic behaviour of some systems,
leading in particular to novel mechanisms of frustration and to
stabilisation of spin-liquid, and maybe orbital liquid states. Spin
and orbital exchange in this situation may become strongly
bond-dependent.  For orbital ordering a straightforward mechanism of
this effect may be understood from Fig.~\ref{FIG:1}: we see that the electron
hopping and the respective exchange for the $e_g$ orbital $|x^2\rangle =
|2x^2-y^2-z^2\rangle$ extended in the $x$-direction is strong for the $ij$-bond
parallel to~$x$, whereas the orbital $|y^2\rangle = |2y^2-x^2-z^2\rangle$ would have
strong overlap with similar orbital in the $y$-bond. If we describe these
doubly-degenerate orbitals by a pseudospin~$\tau$, we would expect an
interaction of the type
\begin{equation}
H = \sum_{\langle ij\rangle_x} \tau_i^x \tau_j^x + \sum_{\langle ij\rangle_y} \tau_i^y \tau_j^y + \sum_{\langle ij\rangle_z} \tau_i^z \tau_j^z\,.
\label{eq:8}
\end{equation}
This model --- the ``compass'' model --- was introduced in~\cite{KugelKhomskii}, Ch.~10,
where it was written: ``Although the interaction within each pair is a
sort of Ising interaction, the overall symmetry is considerably more
complicated. The `cubic' model in~(34) [here~(\ref{eq:8})] has much in common
with the real dipole--dipole interaction, and its properties, even its
ground-state structure, are not yet clear. Pairs arranged along the $z$
axis, for example, would like to align their spins along~$z$, while
pairs arranged along the $x$ axis would like to orient their spins along
this axis. This situation can be described by the classical model of
a lattice of magnetic needles (`compasses'). In contrast with a
one-dimensional system, the properties of two-dimensional and
three-dimensional systems are not understood even qualitatively.''
There was since then a significant theoretical development in this
model, summarised in~\cite{Nussinov}.

A very important version of this model was proposed independently by
Kitaev for the 2d honeycomb lattice, where $xx$, $yy$ and $zz$ interactions
exist on three types of bonds in this lattice~\cite{Kitaev}. In this
situation such model can be solved exactly, and the solution in a
symmetric case is a spin liquid state with very unusual excitations ---
Majorana fermions, which potentially can be even used for quantum
computing, etc. In a very important paper G.~Jackeli and G.~Khaliullin~\cite{Jackeli2}
have shown that this model can be realised in real
TM compounds with honeycomb lattice and with strong spin--orbit
interaction, which exist for example in $4d$ and $5d$ compounds. These
suggestions gave rise to enormous activity in studying such effects,
both experimentally and theoretically, see e.g.~\cite{Trebst,Takagi}.
Experimentally there exist
now several materials which, if not exactly described by the Kitaev
model, are still rather close to it: Li$_2$IrO$_3$, Na$_2$IrO$_3$, RuCl$_3$. To the best of my knowledge, a
good experimental realization of the original compass model of~\cite{KugelKhomskii} in 3d
systems has not been found until now.

Besides using orbital freedom to make bond-dependent exchange and
consequently to enhance frustrations and to stabilise
eventual spin liquid states, one can also think of forming an orbital
analogue of spin liquids --- an orbital, or spin-orbital liquid.
Formally this seems relatively straightforward: we usually describe
orbital ordering using the effective pseudospin description of
orbitals, e.g.\ such that one of two degenerate $e_g$ orbital corresponds
to the pseudospin $\tau^z=+\frac12$, and the other to $\tau^z=-\frac12$. And
intersite interaction leading to orbital ordering is described in
this formalism by the effective ``exchange'' interaction of the type
$J\tau_i\tau_j$, cf.\ the expressions (\ref{eq:5}) and~(\ref{eq:8}). And one can think
that, as in some spin systems, at certain conditions we can have for
orbitals, as for spins, not only some type of ordered states but also
orbital liquid states, analogous to spin liquids. Such ideas were
many times discussed in the literature, and there were even
experimental results interpreted in this way, e.g.\ this mechanism
was suggested to explain the properties of LiNiO$_2$~\cite{Reynaud} ---
see however~\cite{Mostovoy2}. In principle this is indeed a possibility. However
there is one very important factor which makes the situation for
orbitals different from that for spins, and which makes the formation
of such orbital liquid states much more difficult, if not impossible.
The point is that, in contrast to spins which ``live'' more or less
independently, a particular orbital occupation implies a corresponding
distribution of electron charge density, and as such it is
intrinsically strongly coupled with the lattice. And on the lattice the
ions are heavy and behave more or less classically --- or at least the
timescale of corresponding fluctuations is much bigger: the lattice is
``slow''. Consequently one cannot expect to have strong quantum
fluctuations in the orbital sector; in this sense orbitals usually
behave much more classically as compared to spins. As a result
the orbital liquid states are much less probable that spin liquids ---
although one cannot exclude that at certain very specific situations
one can still have orbital or mixed spin-orbital liquids.

\section{Spin-orbit coupling}
Yet another very important factor connected with orbitals, which
strongly influences the properties of most magnetic
materials (and not only those), is the spin--orbit coupling. In contrast with
the spin-orbital effects discussed above, real relativistic spin--orbit
interaction acts everywhere, and it can determine the behaviour of many
materials.

The spin--orbit coupling (SOC)
\begin{equation}
H_{\rm SO} = - \lambda \vec L \cdot \vec S
\end{equation}
couples spins with the charge degrees of freedom, which the orbitals
represent, and because of that it couples spins to the lattice,
leading to such phenomena as magnetic anisotropy (and in most cases
strong magnetoelastic coupling and magnetostriction). Only due to SOC
do we have all these very important characteristics of a particular
magnetic material. Without SOC the spin orientation would be completely
decoupled from the lattice, and the directions in the spin space would have
nothing to do with the real directions in a particular crystal, i.e.\
there would not be any magnetic anisotropy, etc.;\ the spin space would
remain spherically symmetric.

\begin{figure}
\centering
\includegraphics{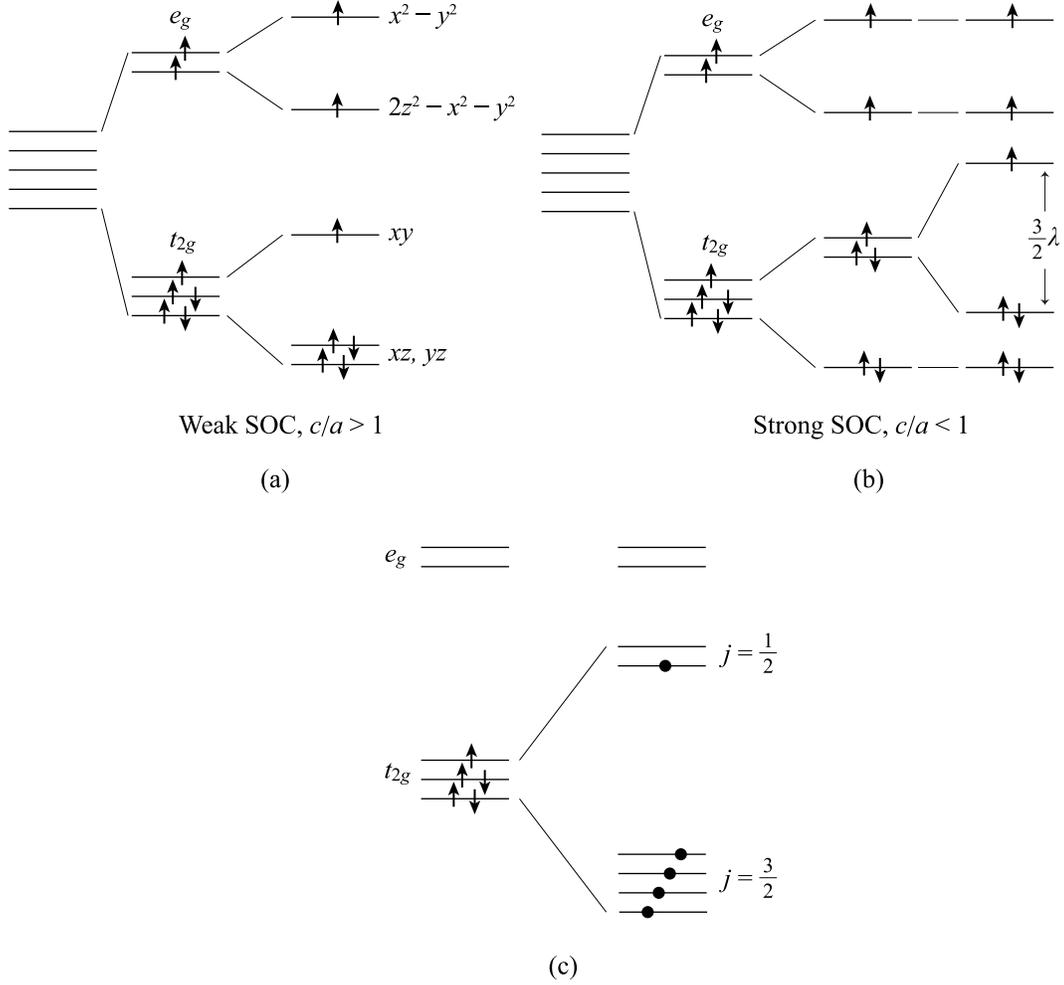}
\caption{\label{FIG:10}\footnotesize Simplified scheme of the modification of energy levels due to spin--orbit
interaction~$\lambda\vec l\cdot\vec S$.
(a)~Energy levels and their filling of transition metals ion with the $d^7$ configuration,
as in Co$^{2+}$, in case of weak spin--orbit coupling, smaller than the Jahn--Teller splitting, which would lead to
Jahn--Teller distortion --- an elongation of $M$O$_6$ octahedra, $\widetilde{c/a}>1$.
(b)~The same for strong spin--orbit coupling, stronger than the eventual Jahn--Teller splitting. It would
lead to the opposite distortion with compressed octahedra, $\widetilde{c/a}<1$.
The situation in real Co$^{2+}$ ions corresponds to the case~(b). Both cases (a) and~(b)
 are obtained in the mean-field picture, valid for bulk systems with strong
intersite interaction. In a more detailed treatment the isolated Co$^{2+}$ ions
have effective spin~$\frac12$ (total spin $S=\frac32$ and effective orbital moment $l_{\rm eff}=1$
give $J_{\rm eff} = \frac12$)~\cite{Abragam}.
(c)~Energy levels of $d$-electrons for very strong spin--orbit coupling (one-electron picture, strictly speaking
applicable for the $jj$ coupling). The electron filling is shown for the $d^5$ configuration,
as in Ru$^{3+}$ or Ir$^{4+}$. One sees that the fifth electron (or one hole) occupies the higher-lying
$j=\frac12$ state which is a Kramers doublet withouth any orbital degeneracy and without the Jahn--Teller effect.}
\end{figure}

All this is very well known and was understood from the very
beginning of the development of the theory of magnetism.  More
specifically, in the typical situation mainly discussed in this
paper, that of TM ions in a cubic crystal field, the orbital moment is
quenched for $e_g$ electrons, and SOC acts on $e_g$ electrons only in
a higher order. However it is not quenched for the $t_{2g}$ triplet; the SOC acts
for those states already in the lowest order, and that is why TM ions
with partially-filled $t_{2g}$ levels typically experience a much stronger
influence of SOC and have much stronger magnetic anisotropy,
magnetostriction etc. Moreover, in many cases the SOC ``interferes'' with
the JT effect, sometimes reversing the sign of lattice distortions.
Thus the JT effect would lead for Co$^{2+}\,(d^7)$ to a tetragonal
elongation, $c/a>1$, but if SOC dominates, one would have not
elongation but compression, $c/a<1$, see Fig.~\ref{FIG:10}(a),~(b)
And experimentally indeed most Co$^{2+}$ systems (CoO, KCoF$_3$)
show such compression, in the case of CoO so strong that it even makes
the magnetic transition at $T_{\rm N}$ a weak first-order transition.  In other
cases, e.g.\ in insulating materials with Ir$^{4+}$, strong SOC leads to
a complete suppression of JT distortion, Fig.~\ref{FIG:10}(c).

These effects become especially interesting and strong for $4d$ and $5d$
systems, in which, first, we typically have low-spin states with
predominant occupation of SOC-active $t_{2g}$ states; and, second, because
SOC becomes much stronger for heavier elements.

Here I want to make one side remark: In most textbooks and other
publications one usually gives the dependence of the SOC on the
atomic number (the charge of the nucleus) $Z$ as $\lambda \sim Z^4$. However
the more accurate treatment presented e.g.\ in another physicists'
``bible'', the course of theoretical physics by Landau and Lifshits,
shows
that actually this dependence should rather be $\lambda \sim Z^2$~\cite{LandauLifshits}. And indeed, a
comparison with the real situation shows that this 
law
fits the experimental data much better. The
conventional dependence~$\sim Z^4$ is just copied from one publication
to the next, in what one calls being ``adopted by repetition'' --- which
does not make it true!

In any case, it is for $4d$ and $5d$ elements that we should expect
strong, and in some case crucial influence of SOC, see e.g.~\cite{Takayama}.  As to the
above-mentioned interplay
between SOC and JT effect, the detailed treatment~\cite{Streltsov4} shows
that it is indeed very strong but not universal, as it strongly depends
on the electron configuration of the respective ions. In most cases
strong SOC suppresses the JT effect. This is not difficult to understand: the JT effect
leads to structural distortions and to occupation by an electron of a
particular orbital with a particular electron density distribution,
actually a particular electric quadrupole moment. Such state is
described by a real wavefunction, e.g.\ for $t_{2g}$ electrons it may be
$|xy\rangle$ or $|xz\rangle$.  However SOC stabilizes the state with a particular
orbital moment which breaks the time-reversal invariance and is described
by a complex wavefunction, e.g.\ $\frac{1}{\sqrt2}(|xz\rangle + i|yz\rangle)$. That is, SOC
and the JT effect typically oppose each other. And this is indeed the case
for low-spin ions with $d^1$, $d^2$, $d^4$ and $d^5$ configurations: here
the JT effect and SOC counteract each other. However the situation
is opposite for the configuration $d^3$~\cite{Streltsov4}: in this case SOC does
not suppress but rather activates, promotes the JT effect. Indeed,
without SOC three $d$-electrons completely fill three $t_{2g}$ levels (with
parallel spins) and there is no orbital degeneracy left.  But for
very strong SOC (when we are close to the $jj$ coupling scheme) the $t_{2g}$ levels,
which can be described by an effective orbital moment $l=1$, are split
into a lower-lying quartet $j=\frac32$ and a higher-lying Kramers doublet
$j=\frac12$, Fig.~\ref{FIG:10}(c). And we now have to put three electrons on this $j=\frac32$ quartet,
which leaves orbital degeneracy. In effect, for the $d^3$
configuration with strong SOC we have the JT effect, and this degeneracy
can be lifted by JT distortions, e.g.\ tetragonal or orthorhombic.
Similarly, one can consider coupling of $t_{2g}$ levels with strong SOC
to trigonal distortions~\cite{Streltsov5}; the results are
qualitatively similar to those of~\cite{Streltsov4}.

These considerations also demonstrate one very interesting situation:
for the $d^5$ configuration, such as is met e.g.\ in Ru$^{3+}$ or Ir$^{4+}$, the strong SOC would
lift orbital degeneracy, and the system would find itself in a nondegenerate
Kramers doublet~$j=\frac12$, Fig.~\ref{FIG:10}(c). Thus the situation seems to be similar to that
of the nondegenerate Hubbard model, Eq.~(\ref{eq:1}).  But the spin--orbit
entanglement in this case strongly modifies all the properties of the
respective materials; it is precisely because of this that honeycomb
systems such as RuCl$_3$ or Na$_2$IrO$_3$ behave as Kitaev systems, as discussed
in Sec.~7.  Thus the strong SOC in $4d$ and $5d$ materials leads to a
profound modification of their state and largely determines their
properties.

Another, potentially much more important consequence of SOC, is that
it can provide a mechanism of formation of nontrivial topological
states --- topological insulators, Weyl semimetals, etc. These
phenomena, very actively studied nowadays, are usually investigated
in systems without strong electron correlations --- in systems such as
Bi$_2$Se$_3$, or TaAs, etc. This is a very intensively developing field
right now, see e.g.\ the reviews~\cite{Hasan,Qi,Ando}. But
these topological properties may exist also in systems with
interacting electrons, in magnetic and other strongly correlated
systems.
The study of this topic still is at an initial stage, but
it already has produced many very interesting results and will definitely
produce much more.

\section{Quantum effects}
In most situation one considers orbitals in TM compounds as classical
objects. This is how we treat the cooperative Jahn--Teller ordering and
the corresponding structural transitions. Basically, this is justified by
the arguments presented in Sec.~7: the orbitals are intrinsically
strongly coupled to the lattice, but the lattice is ``slow'' and
``heavy'', and thus one does not expect strong quantum effects in it.
Nevertheless there are some situations in which one should take
quantum effects in orbital physics very seriously.

The first factor, which is crucial in the treatment of the Jahn--Teller effect in
small systems --- molecules, isolated JT impurities in solids --- is what
is known in the JT field as {\it vibronic effects}. The essence of the JT effect is
that, in contrast to the usual adiabatic approximation, normally used to
treat coupled electron--lattice systems, to each electronic
configurations there corresponds a respective distortion,
\begin{equation}
|\Psi_i\rangle = |\psi_i\rangle |\phi_i\rangle\,,
\label{eq:10}
\end{equation}
where $|\Psi\rangle$ is the total wavefunction --- a vibronic state, and $|\psi_i\rangle$
and $|\phi_i\rangle$ are, correspondingly, wavefunctions of the respective
electronic and lattice (nuclear) configurations. Thus, for example,
to each of the two electronic configurations (orbitals) of two minima of Fig.~\ref{FIG:3}(a),
there corresponds its own distortion. For small systems there will always be
quantum tunnelling between these two states (the two minima of Fig.~\ref{FIG:3}(a)).
These tunnelling processes will lead to significant quantum effects
(quantum fluctuations) even in the ground state of a small JT system.
Here both electronic and lattice
degrees of freedom are equally important, and both have to be treated
quantum-mechanically. This constitutes a big field of {\it vibronic
physics}, a very important part of the treatment of molecules with the
JT effect. If the JT coupling of degenerate electronic states, e.g.\
degenerate orbitals, is not with only one, but say with two
distortion modes, as is the case of $e_g$ electrons (which couple to
doubly-degenerate $E$-type distortions, tetragonal $Q_3$ and orthorhombic
$Q_2$), then the manifold of the degenerate ground states consists not only
of two degenerate minima, as in Fig.~\ref{FIG:3}(a), but is a whole
continuum of degenerate states, the ``trough'' in the
energy surface $E(Q_2, Q_3)$  (the ``Mexican hat'' of
Fig.~\ref{FIG:3}(b)), and the system can freely
rotate along this trough; this has to be treated fully
quantum-mechanically.

One interesting consequence of this full quantum-mechanical treatment
of the combined electron--lattice system is that the nondiagonal matrix
elements of any electron operator are suppressed by the overlap of
nonorthogonal distortion in the vibronic functions~(\ref{eq:10}):
\begin{equation}
\langle\Psi_2|\hat A|\Psi_1\rangle = \langle\psi_2|\hat A|\psi_1\rangle \, \langle\phi_2|\phi_1\rangle\,,
\end{equation}
with the extra factor $\langle\phi_2|\phi_1\rangle < 1$ (often ${\ll}\, 1$).
This reduction is known in the JT field as the {\it Ham's reduction factor}~\cite{Ham,Ham-2};
for the physical community exactly the same effect is known as
{\it polaronic band narrowing}.

Thus, such vibronic, essentially quantum effects, are very important,
and really crucial in the treatment of degenerate situations such as orbital
degeneracy in small systems, molecules and JT impurities in
crystals. However in considering concentrated systems, such as solids
with orbital ordering, we practically always treat the lattice
(quasi)classically. To which extent the ``molecular'' or vibronic quantum
effects might be important for concentrated systems, is an open
question.

But if we ignore the coupling to the lattice, indeed we could expect
quantum effects in orbital sector also in concentrated systems.  There were such suggestions in
the literature~\cite{Khaliullin7,Khaliullin7-2,Ulrich}, especially in
connection with the systems with $t_{2g}$ electrons, for which the JT coupling
to the lattice is weaker. There were claims that some properties of
e.g.\ LaTiO$_3$ and LaVO$_3$ are explained by such quantum fluctuations~\cite{Ulrich}.
However this interpretation is rather questionable, there are
alternative explanations of the observed effects~\cite{Fang,Skoulatos}.

Theoretically there is one very interesting situation, first
considered already in~\cite{KugelKhomskii8,KugelKhomskii}, in which for
the doubly-degenerate orbitals with a particular pattern of electronic
hoppings, with $t_{11}=t_{22}=t$, $t_{12}=0$, the resulting ``Kugel--Khomskii''
model contains both spin and orbital variables~$\vec S$, $\vec\tau$, in the
Heisenberg form: the effective Hamiltonian~(\ref{eq:5}) takes the form
\begin{equation}
H = J\sum_{ij} \Bigl(\vec S_i\cdot\vec S_j + \vec\tau_i \cdot \vec \tau_j + 4 (\vec S_i\cdot\vec S_j)(\vec\tau_i \cdot \vec \tau_j)\Bigr)\,.
\label{eq:12}
\end{equation}
This model has a very high symmetry, SU(4), and it leads to three
gapless excitations (Goldstone modes). It is an extremely quantum model,
which for the 1d case can be solved exactly. This model naturally appears
in TM system with ($M$O$_6$) octahedra with common face~\cite{Kugel,KugelKhomskii3},
and the same model was derived for honeycomb systems with strong SOC
coupling in~\cite{Masahiko}. There it was proposed
that ZrCl$_3$ with honeycomb lattice may be an example of the system
with such interaction, although this suggestion was questioned in~\cite{Ushakov}.

Altogether, the question of possible quantum effects in the orbital
sector of TM compounds is very interesting and largely
still unexplored. As mentioned above, in comparison to spins there is here one important
factor: the coupling of orbitals with the ``heavy'' lattice, which can
suppress quantum effects. But still, there may be situations in which
such quantum effects might be important, up to the possibility of
orbital or spin--orbital liquids.

\section{Orbital effects in metallic systems}
Orbital effects can also be important in metallic systems, although
formally for example the JT effect should work for insulators. Goodenough
in many of his works and in the book~\cite{GoodenoughBook} strongly advocated the
idea that there exist two different thermodynamic states of electrons
in solids --- localised and itinerant. This in itself completely agrees
with the present point of view from which we started this paper. But
as an indicator or the fingerprint allowing us to tell whether a
particular situation corresponds to the one regime or the other, he
suggested to take just the presence or absence of JT distortions: if
in a situation in which formally for a given electron
configuration one would expect orbital degeneracy and JT
distortion but experimentally such distortion is absent, this might be
an indication that we are dealing with itinerant electrons, and vice
versa for localised ones.

Nevertheless there are nowadays many examples which demonstrate that,
despite the absence of a standard JT distortion, in some metallic
systems the orbital effects are still very important. One of the
situations leading to that can be the situation with
orbital-selective localization, discussed in Sec.~5 above. For
example some electrons may be delocalised and may give metallic
conductivity, whereas the other group of electrons behaves as more
localised, and these can display characteristics typical for
localised electrons, including the JT effect.

Another very important situation is when some insulating materials,
with orbital degeneracy and with rather strong JT distortion, become
metallic due to relatively small doping. Here the lattice may
still remain JT-distorted, with the same orbitals playing the main role
in the doped case as well. The most important such example is High-$T_c$
cuprates. For example, the prototype material La$_2$CuO$_4$ is a Mott
insulator with a strong JT effect and with holes occupying one
particular orbital of the $e_g$ doublet, the famous $|x^2-y^2\rangle$ orbital.  And
the same orbitals also remain active in doped systems such as
La$_{2-x}$Sr$_x$CuO$_4$, which are not only metallic but are High-$T_c$
superconductors. Actually for the discoverer of High-$T_c$ cuprates,
Alex Mueller, the fact that Cu is a well-known JT ion was a guiding
principle that led him to search for superconductivity in these
systems --- although in their pure form these factors don't seem to explain
the phenomenon of High-$T_c$ superconductivity in cuprates.   Still,
the very fact that for most of the effects we can work in this case
with a single $(x^2-y^2)$ band, which follows from the JT nature of
Cu$^{2+}$,  demonstrates the general importance of orbital effects in
these systems, as well. There are many arguments that in another class of
High-$T_c$ superconductors, Fe-based superconductors, orbital effects
also play an important role, see e.g.~\cite{Krueger,Fernandes,Yu}.

Yet one more recent development, strongly relying on orbital effects,
is the realization that there exists a special class of materials in
which not so much the Hubbard's~$U$ but the Hund's rule exchange
interaction~$J_{\rm H}$ (operating only in many-orbital
situations) plays a crucial role. These are the so-called Hund's metals~\cite{Haule,Georges}.
All this demonstrates that in many metallic systems the
orbital effects might also be very important and must be taken into
account.

\section{Orbitals in other systems}
\begin{figure}
\centering
\includegraphics{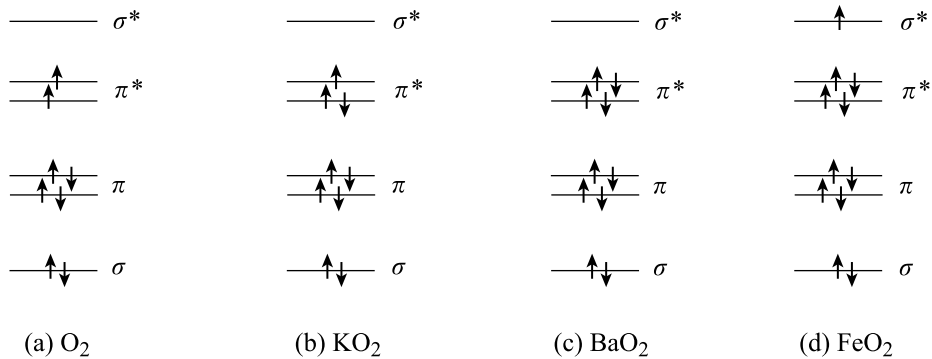}
\caption{\label{FIG:11}\footnotesize Energy levels of O$_2$ dimers and their filling by electrons
for compounds containing oxygen dimers. (a)~O$_2$ molecule.
(b)~Superoxide such as NaO$_2$, KO$_2$.
(c)~Peroxide such as BaO$_2$, MgO$_2$.
(d)~Recently synthesized FeO$_2$~\cite{Streltsov6}.}
\end{figure}

Until now in treating the orbital effects we always had in mind magnetic
systems (transition metal compounds). But in fact the physics
disclosed in studying the orbital effects in TM compounds can be relevant
in many other situations as well. The most straightforward is the
extension of these notions from $d$-electrons to $p$-electrons of, say,
oxygen. It is well known that the oxygen molecules themselves are
magnetic, due to a specific filling of molecular orbitals of those,
Fig.~\ref{FIG:11}(a). Even more interesting effects are observed in superoxides
(or hyperoxides)
such as KO$_2$ and peroxides such as BaO$_2$, Fig.~\ref{FIG:11}(b,c). In these systems the
situation may be indeed very similar to that of $d$-electron systems,
with both magnetic and orbital degrees of freedom strongly coupled~\cite{Solovyev,Wohlfeld}.

But also in completely different systems, such as graphene, including
the twisted bilayer graphene, one meets the situation strongly
resembling that of TM systems with orbital degeneracy. Indeed, due to
a particular lattice geometry leading to a specific band structure, in graphene we
have the situation with extra degeneracy, resulting in
4-fold degenerate states --- two spins and two types of band minima.
And the effective description sometimes used to treat this situation
strongly resembles that of the spin-orbital (Kugel--Khomskii) model~(\ref{eq:5}),
in some cases leading also to the effective SU(4) model, as in~(\ref{eq:12}),
see e.g.~\cite{Yang}.  One
even speaks nowadays, in analogy with spintronics, about ``orbitronics''
(usually having in mind effects caused by spin--orbit coupling) --- see
e.g.~\cite{Vaz}. The experience and the know-how we got in studying
orbital physics in transition metal compounds may be useful in a
broader context of situations in many solids with extra
degeneracy. Thus this section could have a title ``Orbitals everywhere''.

\section{Instead of Conclusions}
Orbital physics, the field initiated by Goodenough in the 1950--1960s
and largely determined by his contributions for many years, has
developed very successfully since then, and it
has served as a basis of realistic description of
magnetic materials --- and not only those. In this short review I have tried to summarise these
very successful developments, and concentrated on several new
directions in orbital physics, which already produced and will
definitely produce many new interesting results. Of course I could
not cover all the aspects of this big field. Thus e.g.\
orbital effects in thin films or at interfaces may also be crucial,
see e.g.~\cite{Konishi,Chakalian}. And they
may be very important in other phenomena as well. In any case, the role played by Goodenough in
this development is indispensable; this entire field carries an
imprint of his great personality.

At the end some personal notes. I came to solid state physics from a
completely different field: my diploma work (master thesis) in the
university was in high-energy physics and in elementary particle
theory. And when after several rather turbulent years I moved to
condensed matter physics, the first thing was to learn the new language.
Rather quickly I came across the book by Goodenough~\cite{GoodenoughBook}, and I
immediately felt that it is a treasure trove of wisdom. But
because of the different background it was initially rather difficult for
me to appreciate and use it. My experience was more or
less the following: when I started to look at a
problem, I would read what Goodenough wrote about it, but very often I
could not understand it. Then I tried to do it my way, and if I
managed to solve the problem, I looked again at what John wrote about
it, and --- ``Aha, that's what he had in mind!''  In fact in all cases
he was right, one only had to be able to understand it. And that is
the message, or rather two messages to the readers:  first, learn a ``foreign''
language; and second, the most important one: Goodenough is
almost always right!  And even if you don't agree with
him on some points (for me this happened only once), it is still extremely
useful and stimulating to know what he did think and write about
the problem.

And the last, just one short  picture. I visited John around 2000,
my only trip to Austin. I stayed there for 2--3 days. Friday
evening we were leaving the institute, it was my last day there. When
we went from John's office to his car, he carried under both arms two
heavy loads of thick volumes of Phys.\ Rev.\ --- older people still
remember how they in the libraries used to make thick bound volumes of three or four
issues of Phys. Rev.
And John explained to me:
``This is my homework for the weekend''. This was about 20 years ago, and
he was already about~80!
It is really incredible how
he continues to keep pace for all these years. It is definitely an
example for all of us --- the example which probably not many
could follow, but at least we have to try. We have a great
role model before us!

\section*{Acknowledgements}
This work was funded by the Deutsche
Forschungsgemeinschaft (DFG, German Research Foundation) -- Project
number 277146847 -- CRC 1238.

\end{document}